\documentclass[10pt,conference]{IEEEtran}
\IEEEoverridecommandlockouts
\usepackage{acronym}
\acrodef{3GPP}{3rd Generation Partnership Project}
\acrodef{6G}{the Sixth Generation}

\acrodef{AoI}{Age of Information}
\acrodef{AoRI}{Age of Request Information}
\acrodef{AoSI}{Age of Sensor Information}

\acrodef{BS}{Base Station}

\acrodef{DRL}{Deep Reinforcement Learning}
\acrodef{RL}{Reinforcement Learning}
\acrodef{DT}{Digital Twin}

\acrodef{FIFO}{First-In-First-Out}

\acrodef{ARVR}[AR/VR]{Virtual and Augmented Reality}

\acrodef{NDT}{Network Digital Twin}
\acrodef{LSTM}{Long Short-Term Memory}
\acrodef{IoT}{Internet of Things}
\acrodef{MEC}{Mobile Edge Computing}

\acrodef{PPO}{Proximal Policy Optimization}
\acrodef{PID}{Proportional–Integral–Derivative}

\acrodef{UE}{User Equipment}

\acrodef{MDP}{Markov Decision Process}

\acrodef{SNR}{Signal-to-Noise Ratio}

\acrodef{SS}{Sensor State}
\acrodef{UR}{User Request}

\usepackage{cite}
 \usepackage{url}
\usepackage{amsmath,amssymb,amsfonts}
\usepackage{algorithmic}
\usepackage{graphicx}
\usepackage{textcomp}
\usepackage{xcolor}

\newcommand{\field}[1]{\mathbb{#1}}
\newcommand{\set}[1]{\mathcal{#1}}

\newcommand{\R}{{\field{R}}}   

\newcommand{\NN}{{\field{N}}}  

\newcommand{\Ks}{{\set{K}}}
\newcommand{\Ss}{{\set{S}}}
\newcommand{\Ns}{{\set{N}}}
\newcommand{\As}{{\set{A}}}

\newcommand{\operator}[1]{\mathrm{#1}}
\newcommand{\laCk}{{\lambda_k^{\operator{(C)}}}}
\newcommand{\laPk}{{\lambda_k^{\operator{(P)}}}}
\newcommand{\laCn}{{\hat{\lambda}_n^{\operator{(C)}}}}
\newcommand{\laPn}{{\hat{\lambda}_n^{\operator{(P)}}}}

\newcommand{\ve}[1]{\boldsymbol{\mathbf{#1}}} 

\newcommand{\vs}{\ve{s}}
\newcommand{\va}{\ve{a}}

\usepackage{graphicx}
\usepackage{url}
\def\BibTeX{{\rm B\kern-.05em{\sc i\kern-.025em b}\kern-.08em
    T\kern-.1667em\lower.7ex\hbox{E}\kern-.125emX}}

\title{
MetaLore: Learning to Orchestrate Communication and Computation for Metaverse Synchronization
}
\author{
\IEEEauthorblockN{Elif Ebru Ohri\IEEEauthorrefmark{1}, 
               Qi Liao\IEEEauthorrefmark{2},
               Anastasios Giovanidis\IEEEauthorrefmark{1},
               Francesca Fossati\IEEEauthorrefmark{1},
               Nour-El-Houda Yellas\IEEEauthorrefmark{3}}
\IEEEauthorblockA{ 
	\IEEEauthorrefmark{1}Sorbonne University, LIP6, Paris, France\\
    \IEEEauthorrefmark{2} Nokia Bell Labs, Stuttgart, Germany\\
	\IEEEauthorrefmark{3}SAMOVAR, Telecom SudParis Institut Polytechnique de Paris, Palaiseau, France\\
E-Mails:
        \IEEEauthorrefmark{1}\{Elif-Ebru.Ohri, Francesca.Fossati, Anastasios.Giovanidis\}\url{@lip6.fr} \\
        \IEEEauthorrefmark{2}\url{qi.liao@nokia-bell-labs.com}, 
	    \IEEEauthorrefmark{3}\url{nour-el-houda.yellas@telecom-sudparis.eu}} 
}

\begin{document}
\maketitle
\begin{abstract}
As\let\thefootnote\relax\footnotetext{This work was supported by a Nokia University Donation.} augmented and virtual reality evolve, achieving seamless synchronization between physical and digital realms remains a critical challenge, especially for real-time applications where delays affect the user experience. This paper presents MetaLore, a Deep Reinforcement Learning (DRL) based framework for joint communication and computational resource allocation in Metaverse or digital twin environments. MetaLore dynamically shares the communication bandwidth and computational resources among sensors and mobile devices to optimize synchronization, while offering high throughput performance. Special treatment is given in satisfying end-to-end delay guarantees. A key contribution is the introduction of two novel Age of Information (AoI) metrics: Age of Request Information (AoRI) and Age of Sensor Information (AoSI) — integrated into the reward function to enhance synchronization quality. An open source simulator has been extended to incorporate and evaluate the approach. The DRL solution is shown to achieve the performance of full-enumeration brute-force solutions by making use of a small, task-oriented observation space of two queue lengths at the network side. This allows the DRL approach the flexibility to effectively and autonomously adapt to dynamic traffic conditions. 

\end{abstract}


\section{Introduction}
Advancements in \ac{ARVR}, high-speed communication, and low-latency edge computing have enabled immersive environments like the Metaverse  \cite{cai2022computedataintensivenetworkskey}, where \acp{DT} act as virtual replicas of real-world objects, continuously receiving real-time data to reflect changes in the physical world \cite{9913665}. 
A key challenge in Metaverse environments is ensuring real-time synchronization between the physical and digital worlds. The digital twin must accurately reflect the changes in the corresponding physical counterpart with minimal delay. This requires instantaneous transmission, processing, and integration of data from sensors and \ac{UE}. However, limited network resources and computation capacity can lead to end-to-end transmission and synchronization delays, causing the virtual replica to fall out of synchronization with the physical entity. Furthermore, variations in service demands, such as the number of connected user devices, sensors, and traffic load, can affect the efficiency of resource allocation. 
The solution lies in dynamically distributing communication bandwidth and computational resources across various data streams—such as user interaction requests and sensor updates, ensuring real-time synchronization. 


In works~\cite{hashash2022towards,Sengendo2024BuildingND,han2022dynamichierarchicalframeworkiotassisted}, 
various models have been proposed for synchronizing digital twins and virtual environments with physical data sources. These include decentralized frameworks based on optimal transport theory~\cite{hashash2022towards}, predictive synchronization in \acp{NDT} using \ac{LSTM} and \ac{PID} control~\cite{Sengendo2024BuildingND}, as well as game-theoretic and evolutionary game approaches for managing synchronization and \ac{IoT} resource allocation ~\cite{han2022dynamichierarchicalframeworkiotassisted}. 
In works ~\cite{8377343,chua2022resource,cheng2024enhancedreinforcementlearningbasedresource}, \ac{DRL} has been used to improve resource management, including computation offloading in MEC~\cite{8377343}, latency minimization for 6G Metaverse services~\cite{chua2022resource}, or training efficiency through \ac{DT}–augmented simulations ~\cite{cheng2024enhancedreinforcementlearningbasedresource}. 


While these works provide valuable contributions, the aspect of synchronizing a device with its environment using periodic sensor data is not adequately integrated. Essentially, a critical gap remains in developing adaptive and dynamic solutions that jointly optimize throughput and end-to-end delay, while keeping an up-to-date version of the device digital twins at the network side through transmission of fresh information from deployed sensors.   
We propose MetaLore, a \ac{DRL}-based framework that dynamically optimizes joint communication and computational resource allocation among devices and sensors, in order to enhance real-time synchronization in Metaverses. Our novel contributions are as follows:
\begin{itemize}
\item \emph{Multi-objective and queuing-aware optimization}: The framework formulates resource allocation among sensors and devices as a multi-objective problem, balancing end-to-end latency, synchronization accuracy, and system throughput, using observations from queues that hold incoming serving packets at the network side. 
\item \emph{Novel \ac{AoI} metrics}: We introduce \ac{AoRI} and \ac{AoSI} to quantify data freshness, ensuring synchronization-aware decision-making. 
\item \emph{Adaptive learning with an efficient, lightweight model}: MetaLore leverages \ac{DRL}, specifically the \ac{PPO} algorithm \cite{schulman2017ppo}, to dynamically split resources among sensors and devices and adapt to traffic evolution. To ensure scalability and efficiency, it utilizes a task-specific observation space based solely on computational queue lengths.
\item \emph{A custom simulation environment is built} to simulate and test the DRL solution. It includes dynamic traffic with user movement, dynamic packet generation and queuing as well as twinning synchronization. It is built upon and extends  \texttt{mobile-env} \cite{schneider2022mobileenv}.
\end{itemize}

The rest of the paper is organized as follows: Section \ref{sec:SysModel} introduces the system model, while Section \ref{sec:ProbForm} presents the problem formulation and proposed \ac{DRL}-based solution. Section \ref{sec:Sim} details the simulation setup, numerical results, and key findings. Finally, Section \ref{sec:Concl} concludes with remarks and future directions.

\section{System Model}\label{sec:SysModel}

\subsection{Environment and System Components}
We model a physical region as a segment of the real world, 
divided into non-overlapping areas, each area 
linked to a distinct decentralized sub-metaverse. Each sub-metaverse 
represents a digital counterpart of a physical area. For each sub-metaverse, a \ac{BS} 
and an \ac{MEC} server 
are assigned.
The \ac{BS} provides communication resources, offering bandwidth for data transfer from connected devices, while the \ac{MEC} server acts as a computational hub, enabling real-time processing of incoming requests and environmental updates. We begin with a single sub-metaverse scenario for scalability, but the model can easily be extended to multiple sub-metaverse scenarios, where each operates independently while following the same principles. 

Each sub-metaverse manages a set of $K$ \acp{UE}, denoted by \( \Ks \), and a set of $N$ sensors, denoted by \( \Ns \).
Each \ac{UE} \( k \in \Ks\) represents a mobile device operating within the coverage area of a \ac{BS}, interacting with the sub-metaverse by generating service requests and device updates. 
Each sensor \(n \in\Ns\) is stationary and distributed across the coverage area, continuously monitoring the physical environment. These sensors capture real-time data on environmental conditions or device status within their sensing range and transmit updates at fixed intervals to maintain up-to-date digital twins.

Subsequently, service requests of \ac{UE} and sensor updates are characterized by the size of data packets\footnote{Here, a \lq\lq packet" may refer to either a single data packet or a large aggregated packet (e.g., a bundle of multiple smaller packets) associated with a request with a size smaller or equal to the largest protocol data unit (PDU).} for transmission and the corresponding processing time for computation. 
In a discrete-time system, each mobile device $k\in\Ks$ generates update requests at random time intervals following a Bernoulli distribution, where the probability of generating a request in a time slot is \( p_k \). Namely, the number of requests generated by \ac{UE} $k$ at time $t$ follows  $X_k(t)\sim \operator{Bernoulli}(p_k)$. The size of the requested packet \( \beta_k \) for communication follows a Poisson distribution with mean \( \laCk \), while the computational load \( \psi_k \) for processing the request on the \ac{MEC} server follows a  Poisson distribution with mean \( \laPk \). 
In contrast, sensor updates are periodic. The number of updates of sensor $n$ at time t yields $\hat{X}_n(t) = \delta_{t~\operator{mod}~T_n, 0}$, where $\delta_{i,j}$ is the Kronecker delta function and $T_n$ is a constant update interval. Each update of sensor $n\in\Ns$ has a data size $\hat{\beta}_n(t)\sim\operator{Poisson}\left(\laCn\right)$. The computational load for processing each sensor update follows \( \hat{\psi}_n \sim\operator{Poisson}\left(\laPn\right)\). Note that, to distinguish the notations for metrics related to UEs and sensors, we place a hat over all sensor-related notations. 


\subsection{Queueing System}
\begin{figure}[t]
    \centering \includegraphics[width=0.95\columnwidth,clip,trim=0.5pt 0 0.5pt 0]{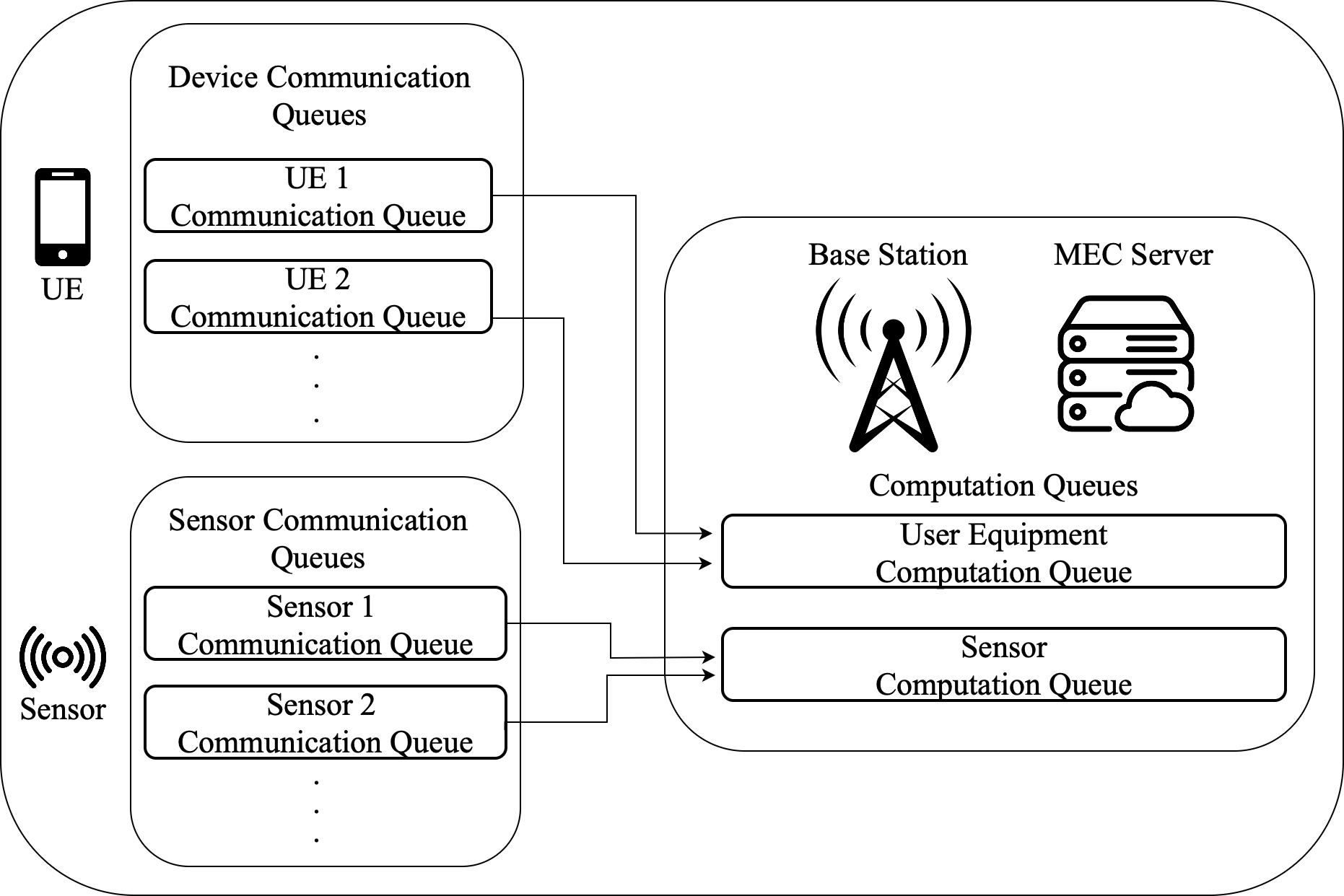}
    \begin{sloppypar}\caption{Illustration of the queuing mechanism.}\end{sloppypar}
    \label{fig:device_queue}
\end{figure}

Inside a sub-metaverse, a queuing system regulates the flow of data between various system components throughout transmission and processing, as shown in Fig.~\ref{fig:device_queue}. The queueing system follows a \ac{FIFO} principle and it handles packet flow between \ac{UE} and \ac{BS}, between sensor and \ac{BS}, as well as between \ac{BS} and \ac{MEC} server. This mechanism facilitates tracking requests and prevents packet loss even in scenarios with high data traffic or network congestion.

At the device level, each \ac{UE} and sensor is equipped with a dedicated \emph{communication queue} inside the device to manage the scheduling and transmission of service requests to the \ac{BS}. Initially, if \ac{UE} $k$ generates a request entering the device queue, i.e., $X_k(t)=1$, it demands bandwidth based on the data size of the generated request $\beta_k(t)$. Similarly, sensor $n$ at time $t$ requires bandwidth depending on $\hat{X}_n(t)\hat{\beta}_n(t)$. 
Device communication queues temporarily buffer data packets until they are successfully transmitted to the \ac{BS}. 

At the \ac{BS}, incoming data from devices is categorized into two distinct \emph{computation queues}: one for UE-generated traffic and another for sensor data. These computation queues temporarily buffer data packets until they are allocated computational resources by the \ac{MEC} server. Let $l(t)$ and $\hat{l}(t)$ denote the queue lengths of accumulated traffic from UEs and sensors at time $t$, respectively. Within the server, each service request in the computation queues awaits processing, requiring computational resource according to its processing load, $\psi_k(t)$ for \ac{UE} $k$ and $\hat{\psi}_n(t)$ for sensor $n$, respectively.


\subsection{System Delay and Synchronization}
The modeling of delay is another fundamental aspect of the system which quantifies the \ac{AoI} provided by the \ac{UE} and sensors. It emphasizes the efficiency of data transmission and synchronization of user requests with environmental updates from sensors. The system considers two types of delay:

    \subsubsection{Age-of-Request-Information (\ac{AoRI})
    } refers to the end-to-end service delay, i.e., the time interval between the generation of the $i$-th  request from \ac{UE} $k\in\Ks$ and the moment it is fully served. It is expressed as:
    \begin{equation}
    AoRI_{k, i} = T^{(\operator{Q})}_{k, i}\!+\!T^{(\operator{T})}_{k,i}\!+\!T^{(\operator{B})}_{k,i}\!+\!T^{(\operator{P})}_{k,i}, \mbox{for } i\in\NN_{+},
    \label{eqn:E_delay}
    \end{equation}
    where, \( T^{(\operator{Q})}_{k, i} \) represents the time that request $i$ spent in the transmission queue of user device $k$ before transmission starts, \(  T^{(\operator{T})}_{k, i} \) denotes the transmission time from device to the \ac{BS}, \( T^{(\operator{B})}_{k, i} \) represents the time spent at the \ac{BS} until processing starts and \( T^{(\operator{P})}_{k, i} \) represents the processing time at the \ac{MEC} server. 

    \subsubsection{Age-of-Sensor-Information (\ac{AoSI})}
\begin{figure}
            \centering \includegraphics[width=0.9\linewidth]{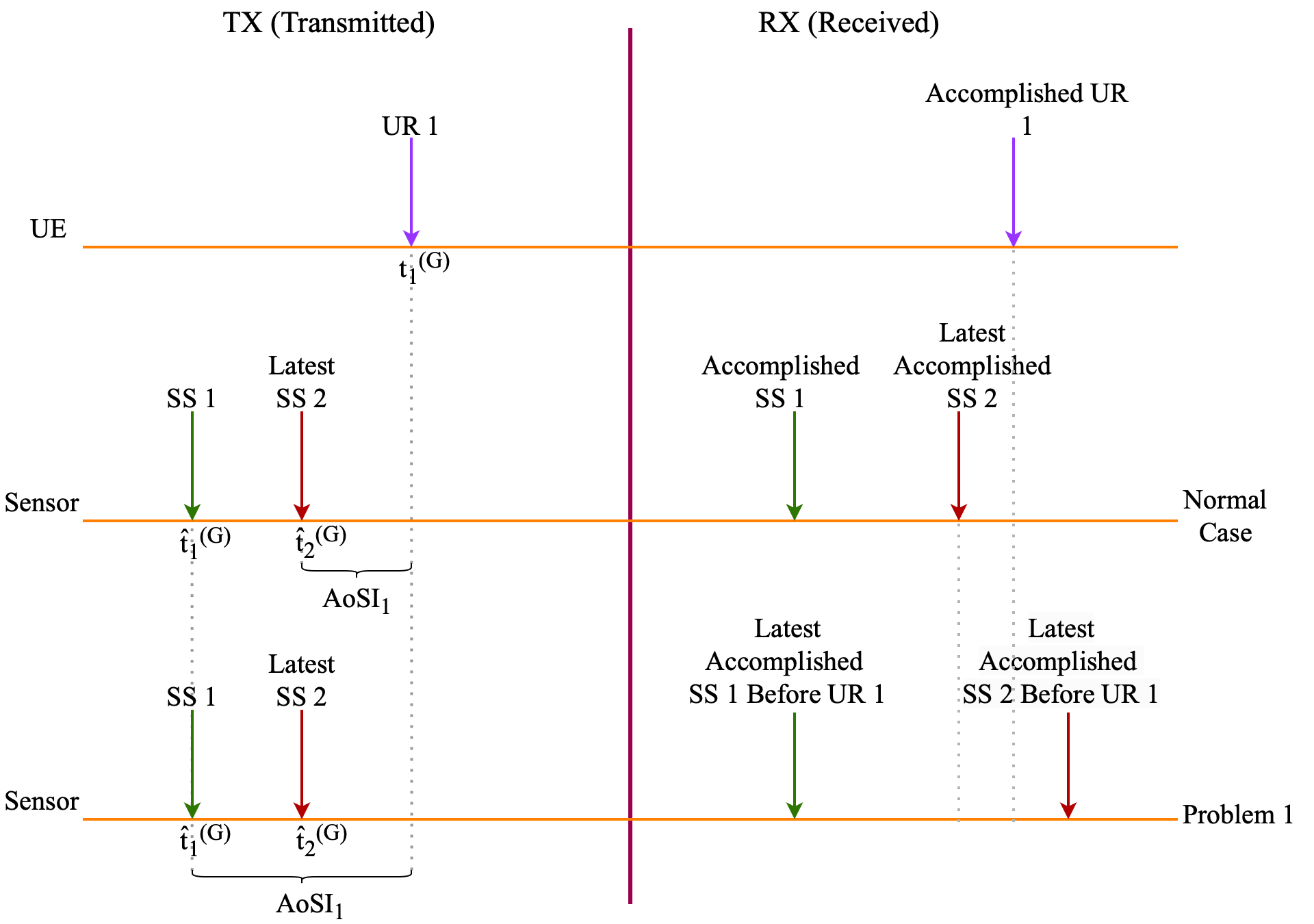}
            \caption{Illustration of late arrival of new sensor state: a newer sensor update arrives after the completion of a user request.
            }
            \label{fig:delay_computation1}
        \end{figure}

    quantifies the synchronization delay, i.e., the time lag between the generation time of the $i$-th service request from \ac{UE} $k$, denoted by \( t_{k,i}^{(\operator{G})} \),  and the generation time of the most recent processed environmental update from the sensors covering \ac{UE} $k$'s location, denoted by $\hat{t}_{n^{\ast}(k), j^{\ast}(i)}^{(\operator{G})}$. Note that $n^{\ast}(k)$ denotes the selected sensor that covers \ac{UE} $k$'s location, while $j^{\ast}(i)$ denotes the index of the most recent processed updates from the selected sensor, before request $i$ has been processed in \ac{MEC} (examples in Fig.~\ref{fig:delay_computation1}). Thus, the synchronization delay is defined as:
    \begin{equation}
    AoSI_{k,i} = t_{k,i}^{(\operator{G})} - \hat{t}_{n^{\ast}(k), j^{\ast}(i)}^{(\operator{G})}, \mbox{for } k\in\Ks,  i\in\NN_{+}.
    \label{eqn:S_delay}
    \end{equation}
    The above expression quantifies how outdated the sensor data is when serving a user request. This synchronization delay computation may encounter two critical cases:
    
    $\bullet$  \textit{Late arrival of sensor update:}  
        As illustrated in Fig.~\ref{fig:delay_computation1}, sensor states (SSs) $j=1, 2$ are generated before user request (UR) $i=1$. For brevity, let us omit the \ac{UE} index $k$ and sensor index $n$ in this example, but only use the \ac{UE} request index $i$ and sensor update index $j$. The latest sensor state before the generation of request $i=1$ is $j=2$. In the normal case, the update $j=2$ is successfully processed before the processing time of request $i=1$. Therefore, $j^{\ast}(i)=2$, and the synchronization delay is 
        $AoSI_{1} = t_{1}^{(\operator{G})}-\hat{t}_{2}^{(\operator{G})}$. 
        However, in Problem 1, 
        sensor state $j=2$ is processed only after the user request $1$ is fulfilled. Consequently,
        $$AoSI_{1} = t_{1}^{(\operator{G})} - \hat{t}_1^{(\operator{G})}.$$
        \vspace{-2ex}
        
\begin{figure}
            \centering \includegraphics[width=0.9\linewidth]{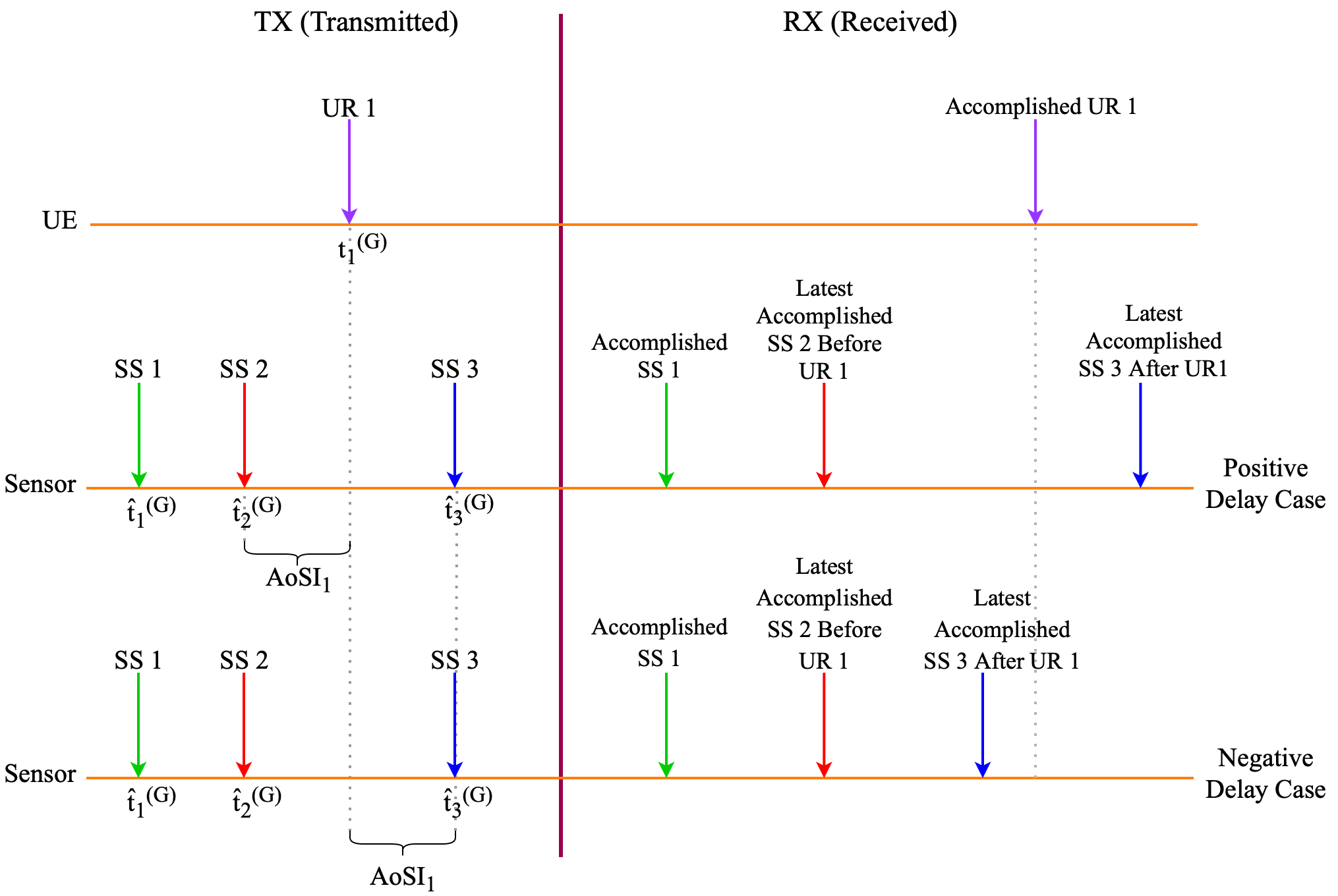}
            \caption{Illustration of the early arrival of new sensor state: a sensor update generated after the user request is processed earlier, resulting in a negative synchronization delay.
            }
            \label{fig:delay_computation2}
            \vspace{-2ex}
        \end{figure}

    $\bullet$    \textit{Early arrival of sensor update:}  
        As depicted in Fig.~\ref{fig:delay_computation2}, sensor state $j =3$ is generated after user request $i=1$. If resource allocation prioritizes sensor data processing, sensor states generated after a user request, such as sensor state 3, may be processed before the completion of the service of the earlier user request 1. This results in a negative synchronization delay. To accurately capture the sensor information and ensure a non-negative measure of delay, the synchronization delay is computed as the absolute time difference between the generation time of the user request 1 and the generation time of the most recently served sensor update of state 3:
        $$
        AoSI_1= \left|t_{1}^{(\operator{G})} - \hat{t}_3^{(\operator{G})}\right|.$$
\vspace{-2ex}
\section{Problem Formulation and Learning Approach}\label{sec:ProbForm}

\subsection{Markov Decision Process}
We model the joint communication and computational resource
allocation problem as a \ac{MDP}, defined by a tuple $\left\{\Ss, \As, P(\cdot), r(\cdot), \gamma\right\}$, where $\Ss$ denotes the state space, $\As$ is the action space, $P:\Ss\times\As\times\Ss\to [0,1]$ indicates the transition dynamics, 
the reward function is represented by $r:\Ss\times\As\to\R$, reflecting the synchronization and end-to-end service latency for all accomplished \ac{UE} service requests, and $\gamma\in[0,1]$ is the discount factor.
The transition from one state to the next is determined by how the communication and computation queues evolve and how successfully the requests are processed, based on the chosen resource allocation at each time step.

    \subsubsection{State Space}
    At time step $t$, the system observes the state $\vs(t)$ which is represented by the size of two computation queues at the \ac{BS}, as illustrated in Fig.~\ref{fig:device_queue}. Specifically, the state includes the two computational queue lengths of accumulated UE- and sensor-generated tasks,
    \begin{equation}
    \vs(t) = \left[l(t), \hat{l}(t)\right] \in\Ss\subseteq\R^2.
    \label{eqn:state_space}
    \end{equation}

    \subsubsection{Actions}
At decision time~$t$, the agent selects an action $\va(t)$ that determines the allocation ratios of communication and computational (processing) resources to the \acp{UE}, denoted by $\rho^{\operatorname{(C)}}$ and $\rho^{\operatorname{(P)}}$, respectively. The remaining resources are allocated to the sensors, given by $\hat{\rho}^{\operatorname{(x)}} = 1 - \rho^{\operatorname{(x)}}$, for $x \in \left\{\operatorname{C}, \operatorname{P}\right\}$. The action space $\mathcal{A}$ is continuous and defined as:
    \begin{equation}
    \va(t) = \left[\rho^{\operator{(C)}}(t), \rho^\operator{(P)}(t)\right] \in \As:= [0,1] \times [0,1].
    \label{eqn:action_space}
    \end{equation}

    \subsubsection{Rewards}
    At each time step $t$, multiple \ac{UE} service requests can be generated and potentially fulfilled, depending on the system state and the chosen resource allocation. The objective is to concurrently minimize the synchronization delay and end-to-end service delay for each \ac{UE} request while maximizing the aggregate \ac{UE} throughput. The reward function is designed accordingly: 
    Let $r^{(\operatorname{S})}(t)$ denote the synchronization reward and $p^{(\operatorname{E})}(t)\leq 0$ the penalty for end-to-end service delay. The total reward at time $t$ is given by:
    \begin{equation}
    r(t) = r^{(\operator{S})}(t) + p^{(\operator{E})}(t).
    \label{eqn:total_reward}
    \end{equation}

    $\bullet$ \textit{Synchronization Reward}:
    At each time step $t$, the synchronization reward is calculated only for the \ac{UE} service requests that are successfully completed at that specific time step, denoted by the set $\mathcal{I}_k(t)$ for each UE $k$ 
    \begin{equation}
    r^{(\operator{S})}(t) = \sum_{k \in \mathcal{K}} \sum_{i \in \mathcal{I}_k(t)} c_1 \cdot \eta^{AoSI_{k, i}}, 
    \label{eqn:s_reward}
    \end{equation}
    where \( c_1 \) is the constant base reward for synchronization, $\mathcal{I}_{k(t)}$ is the set of existing requests from user $k$ at time $t$, and \( \eta \in (0, 1) \) is the delay discount factor for synchronization. This expression gives $c_1$ reward if the served request has $AoSI = 0$ but discounts this base reward as the sync gap becomes larger. 

    $\bullet$ \textit{Delayed End-to-end Service Penalty}: 
    At each time step~$t$, a penalty is applied to requests completed at~$t$ whose end-to-end delay exceeds the threshold $D_{\text{th}}$; no penalty is incurred for timely completions. The total delay penalty at time~$t$ is defined as:
    \begin{equation}
    p^{(\operator{E})}(t) = \sum_{k \in \mathcal{K}} \sum_{i \in \mathcal{I}_k(t)} c_2 \cdot 
    \mathbf{1}_{\{
    AoRI_{k, i} > D_{\text{th}}\}},
    \label{eqn:penalty}
    \end{equation}
    where \( c_2 \) is the base penalty per delayed \ac{UE} request.
    
    


\subsection{Problem Formulation}
The objective of the reinforcement learning agent is to learn an optimal policy \(\pi: \mathcal{S} \rightarrow \mathcal{A}\) that decides the proportion of the communication and computational resources allocated for \acp{UE}, leaving the remaining for the sensors. 
Formally, the RL optimization problem is expressed as:
\begin{equation}
\max_{\pi} \; \mathbb{E}_{\pi} \left[ \sum_{t=0}^{T} \gamma^t r(\vs(t), \va(t)) \right], \quad \text{subject to } \va(t) \in \As.
\label{eq:rl_objective}
\end{equation}

This formulation supports learning a dynamic resource allocation policy that responds to time-varying system and heterogeneous service demands. 
 Its goals are to ensure timely synchronization of \ac{UE} tasks with fresh sensor data, minimize end-to-end delay, and maintain high throughput. To achieve this, the policy must balance these competing objectives in real time for optimal overall system performance.
 
Despite its structured formulation, the problem remains challenging due to its inherently multi-objective nature, where synchronization accuracy, end-to-end latency, and throughput are interdependent and often conflicting. For instance, allocating more communication bandwidth or computation capacity to sensors may improve synchronization freshness but at the cost of increased delay or reduced throughput for \ac{UE} requests. Conversely, prioritizing \ac{UE} tasks may degrade the alignment between user actions and sensor updates. These trade-offs should be accounted for by the solution. 

\subsection{Deep Reinforcement Learning Algorithm}
To solve the formulated reinforcement learning problem, we adopt the popular on-policy \ac{PPO} algorithm \cite{schulman2017ppo}, a widely used actor-critic method which improves policy stability by limiting updates through a clipped surrogate objective. PPO is particularly well suited to our problem setting due to its robust performance in continuous action spaces, as well as its stable learning behavior for the constrained, multi-objective nature of our scenario.

Let \( \pi_\theta \) denote the current policy with parameters $\theta$, \( \pi_{\theta_{\text{old}}} \) the previous policy, and \( \hat{A}_t \) the estimated advantage at time step \( t \). For brevity, we use the subscript \( t \) notation throughout this subsection. The probability ratio is defined as:
\[
r'_t(\theta) = \frac{\pi_\theta(\va_t|\vs_t)}{\pi_{\theta_{\text{old}}}(\va_t|\vs_t)}.
\]

Given a small positive clipping parameter $\epsilon$, the clipped policy loss is:
\[
\mathcal{L}^{\text{CLIP}}(\theta) = \mathbb{E}_t \left[ \min \left( r'_t(\theta) \hat{A}_t, \, \text{clip}(r'_t(\theta), 1 - \epsilon, 1 + \epsilon)\hat{A}_t \right) \right].
\]

The value function loss and entropy bonus are given by:
\[
\mathcal{L}^{\text{VF}}(\theta)=\mathbb{E}_t\!\big[(V_\theta(\vs_t)-\hat V_t)^2\big],\;
\mathcal{L}^{\text{S}}(\theta)=\mathbb{E}_t\!\big[\mathcal{H}(\pi_\theta(\cdot \mid \vs_t))\big].
\]

The total PPO objective combines these components:
\[
\mathcal{L}^{\text{PPO}}(\theta) = \mathcal{L}^{\text{CLIP}}(\theta) - c_1 \mathcal{L}^{\text{VF}}(\theta) + c_2 \mathcal{L}^{\text{S}}(\theta).
\]
PPO alternates between data collection using \( \pi_\theta \) and several epochs of gradient ascent on \( \mathcal{L}^{\text{PPO}} \).



\section{Simulation Setup and Results}\label{sec:Sim}
To evaluate the effectiveness of the proposed \ac{DRL}-based resource allocation framework, we developed a custom simulation environment built on \texttt{mobile-env} \cite{schneider2022mobileenv}, extended to model sub-metaverse dynamics including dynamic packet generation for \ac{UE} and sensor, queuing, and request synchronization. The environment is implemented with \texttt{Gymnasium} \cite{towers2024gymnasium}, and training is performed using the \ac{PPO} algorithm from \texttt{Stable-Baselines3} \cite{stable-baselines3}. The code is publicly available at: \url{https://github.com/elifohri/MetaLore-simulator}.

\subsection{System Parameters and Simulation Configuration}
We simulate a single sub-metaverse system comprising one \ac{BS}, one \ac{MEC} server, 15 static sensors, and 10 mobile \acp{UE}. The \ac{BS} operates with a total communication bandwidth \(R^{(C)} = 100~\mathrm{MHz}\), a carrier frequency of 3.5~GHz, and a transmission power of 40~dBm. It is placed at a height of 40 meters and is equipped with a processing capacity \(R^{(P)}\) of 100 units. 
Each \ac{UE} moves at an average velocity of 1.5~m/s. 
An antenna height of 1.5 meters used for both the \ac{UE} and sensors. At each timestep, a \ac{UE} \(k \in \mathcal{K}\) generates a service request with probability \(p_k = 0.7\), modeled as Bernoulli process described in Section~\ref{sec:SysModel}. Communication and computational loads (\(\beta_k\), \(\psi_k\)) modeled as Poisson random variables with means \(\lambda^{(C)}_k = 50\) and \(\lambda^{(P)}_k = 5\), $\forall k$. Each sensor \(n \in \mathcal{N}\) is placed at a height of 1.5 meters and generates updates with higher resource demands with time interval $T_n=1$: communication and computational load follows Poisson distributions with means \(\hat{\lambda}^{(C)}_n = 70\) and \(\hat{\lambda}^{(P)}_n = 7\), respectively, $\forall n$.

\subsection{Static Resource Allocation Baseline}
To establish a comparative benchmark, we implement a static resource allocation baseline using an exhaustive grid search of computation $\times$ communication resources $\{0, 0.1,\ldots,1\}^2$. Each fixed allocation is evaluated over 100 episodes of 100 time steps with varying random seeds to ensure statistical robustness. 

\subsection{Learning-Based Dynamic Resource Allocation}
In the proposed \ac{DRL} framework, the reward function includes four key parameters introduced in Section~\ref{sec:ProbForm}: base synchronization reward \(c_1\), synchronization delay discount factor \(\eta\), base penalty for delayed \ac{UE} requests \(c_2\), and end-to-end delay threshold \(D_{\mathrm{th}}\). 
For experiments, we fix \(c_1 = 10\) and \(D_{\mathrm{th}} = 2\) time steps. The focus is on tuning \(\eta\) and \(c_2\). The discount factor \(\eta \in (0,1)\) controls the sensitivity in synchronization delay, the lower the $\eta$ the less valuable the received packet if it uses stale sensor information. The penalty  \(c_2\) discourages excessive \emph{end-to-end service delay} 
via negative rewards. These parameters also  indirectly affect throughput. 

Training proceeds in two phases. Phase 1 fixes \(c_2 = 0\) and varies \(\eta \in \{0.9, 0.8, \ldots, 0.4\}\) to assess the impacts on synchronization delay and total \ac{UE} throughput. Phase two introduces penalty \(c_2 = -1\) for requests exceeding \(D_{\mathrm{th}}\), maintaining the same range of \(\eta\), to evaluate joint synchronization-latency performance. This systematic tuning reveals optimal reward settings for the desired trade-offs.



\subsection{Results and Analysis}

\begin{figure}[t]
    \centering
    \includegraphics[width=0.95\linewidth]{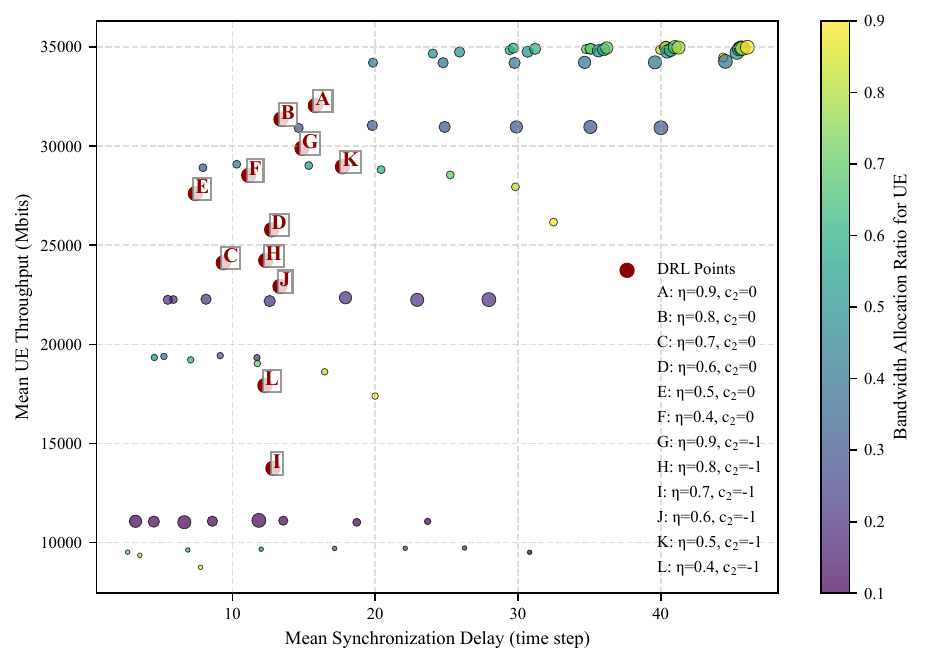}
    \caption{Throughput vs mean synchronization delay (AoSI). Red circular markers labeled A–L denote the performance of proposed DRL approach under different hyperparameters.  Colormapped dots represent baseline results with fixed bandwidth splits, and their sizes indicate the computational resource allocation. 
    }
    \label{fig:throughput_vs_sync}
    \vspace{-2ex}
\end{figure}

\begin{figure}[t]
 \centering
 \includegraphics[width=0.95\linewidth]{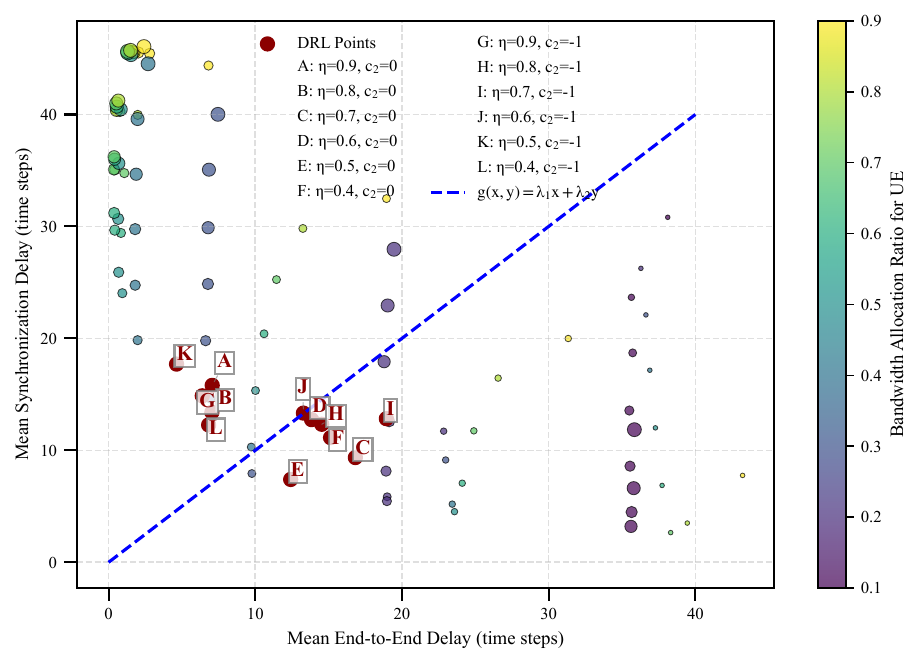}
 \caption{Mean synchronization delay (AoSI) vs mean end-to-end service delay (AoRI). 
 }
 \label{fig:synch_vs_e2e}
\end{figure}


\begin{table}[t!]
\begin{tabular}{|c|c|c|c|}
\hline
\textbf{(\(\rho^{(C)}\), \(\rho^{(P)}\))} & \textbf{Sync. Delay} & \textbf{E2E Delay} & \textbf{Traffic Served \%} \\
\hline
(0.3, 0.3) & 7.93 $\pm 1.41$ & 9.77 $\pm 1.23$ & 82.61 \\ 
(0.3, 0.4) & 14.63 $\pm 1.43$ & 6.94 $\pm 1.53$ & 88.35 \\ 
(0.4, 0.4) & 1.98 $\pm 0.50$ & 19.84 $\pm 1.25$ & 97.76 \\ 
\hline
\hline
DRL \textbf{($\eta$, $c_2$)} &  \textbf{Sync. Delay} & \textbf{E2E Delay} & \textbf{Traffic Served \%} \\ 
\hline
A ($0.9$, $0$) & 15.81 $\pm$ 2.71 & 7.07 $\pm$ 1.46 & 91.57 \\ 
B ($0.8$, $0$) & 13.39 $\pm$ 1.42 & 7.04 $\pm$ 0.92 & 89.60 \\ 
E ($0.5$, $0$) & 7.40 $\pm$ 1.40 & 12.42 $\pm$ 1.94 & 78.88 \\ 
G ($0.9$, $-1$) & 14.87 $\pm$ 2.11 & 6.39 $\pm$ 1.83 & 85.42 \\ 
K ($0.5$, $-1$) & 17.70 $\pm$ 1.90 & 4.64 $\pm$ 1.23 & 82.76 \\ 
\hline
\end{tabular}
\vspace{1mm}
\caption{Mean $\pm$ standard deviation of selected fixed and DRL policies. Delays are in \textit{time steps}; throughput is given as percentage of the maximum (35,000 Mbits).}
\label{tab:drl_points}
\end{table}



Fig.~\ref{fig:throughput_vs_sync} presents a clear trade-off between synchronization delay and \ac{UE} throughput for both static and DRL-based resource allocation strategies, across various discount factors $\eta$ and penalties $c_2$. 
Static policies that achieve high \ac{UE} throughput exhibit increased synchronization delays, suggesting that a larger fraction of resources are allocated to serving \ac{UE} request completion rather than ensuring timely updates from sensors. Conversely, policies that minimize synchronization delay often achieve reduced \ac{UE} throughput, highlighting the competing objectives. 
In contrast, DRL policies achieve superior trade-offs, with several points—specifically A (\(\eta=0.9\)), B (\(0.8\)), and E (\(0.5\)). They are placed on the Pareto frontier, i.e. there is no other fix-split allocation exceeding them, and the tangent of the frontier is visually around $45^o$ meaning that they achieve a good trade-off, where both synchronization delay and throughput are jointly optimized. 
In particular, points A, B, and E are located on the upper-left boundary of the plot, indicating that they dominate other fixed allocation configurations by offering the best available trade-offs, i.e., minimizing synchronization delay without sacrificing throughput, or maximizing throughput without incurring excessive synchronization penalties. 
When examining the behavior of DRL policies with lower discount factors, specifically points C (\(\eta=0.7\)) and D (\(\eta=0.6\)), they exhibit a drop in \ac{UE} throughput without a proportional improvement in synchronization delay. This behavior implies that the reward function for such low $\eta$ values allocates an unnecessarily high amount of resources to sensors, while neglecting the devices. 

Fig.~\ref{fig:synch_vs_e2e}, illustrates the policy achievable trade-off between end-to-end service delay (AoRI) and the synchronization delay (AoSI) for both DRL-based policies and static allocation policies. Points A (0.9), B (0.8) and E (0.5) perform very well, with E being again Pareto optimal. When an end-to-end service penalty is introduced ($c_2=-1$), the DRL agent adjusts to reduce service delays.  Points L ($\eta = 0.4$) and K ($\eta = 0.5$), corresponding to nonzero penalty cases, achieve similar end-to-end delays compared to the no-penalty points A, B and E while maintaining low synchronization delay. A glimpse in Fig.~\ref{fig:throughput_vs_sync} shows, however, that they compromise in the \ac{UE} throughput. In Fig.~\ref{fig:synch_vs_e2e}, the most desirable solutions lie in the lower-left corner, where most DRL points lie. Static split allocations can achieve Pareto optimal points but only at extremes or through exhaustive search, e.g., the (0.3, 0.3) split. Since real systems can't brute-force all options, such a solution is impractical. In contrast, the DRL agent learns optimal resource management from state observations alone, making it suitable for real-time deployment. 


Table~\ref{tab:drl_points} presents a detailed numerical comparison between the performance metrics of three fixed-split baselines and the Pareto-optimal points identified by the DRL policy. The baseline splits $(0.3, 0.3)$ and $(0.3, 0.4)$ are obtained via exhaustive search, while $(0.4, 0.4)$ is heuristically derived based on the ratio of number of UEs to sensors. The results demonstrate that the DRL policies can effectively learn to operate near-optimal split configurations. In particular, DRL points A, B, and G closely align with the $(0.3, 0.4)$ baseline, prioritizing low AoRI (end-to-end delay) while achieving improved throughput. DRL point E approximates the $(0.3, 0.3)$ baseline, favoring lower synchronization delay at the cost of increased end-to-end delay. In contrast, the heuristic split $(0.4, 0.4)$, though achieving high throughput, suffers from significantly higher end-to-end delay, highlighting the limitations of naive proportional allocation.
An operator may choose the appropriate values for the reward $(\eta, c_2)$ in order to guide the solution towards the desirable trade-off, depending on service requirements. 

In addition to its performance, the DRL approach benefits from a lightweight observation design. By using a small, task-oriented observation space, based only on computational queue lengths, DRL approach becomes flexible and adaptable to varying traffic conditions. When these evolve, e.g., when the number of devices changes, or when the packet generation and service requirement statistics change, the DRL solution will use the updated queue length observations as states to update the bandwidth-computation allocation. The resulting delay and throughput metrics will be equivalent to the best resource split corresponding to the new traffic scenario, but without having to exhaustively test and evaluate every possible split configuration once $(\eta, c_2)$ are empirically tuned.

\section{Conclusion and Discussion}\label{sec:Concl}
In this work we proposed a DRL-based framework for dynamic resource allocation in a metaverse environment related to a single cell. The DRL policy can optimize multiple objectives, minimizing both synchronization and end-to-end service delays, while maximizing throughput, by learning to dynamically split communication and computation resources between sensors and mobile users, using queue-length information at the base station.  
Future work will extend this approach to multi-base-station scenarios, introducing new challenges such as inter-BS synchronization, resource sharing, and user association across interconnected sub-metaverses. 

\bibliographystyle{IEEEtran}
\bibliography{references}

\end{document}